\documentclass[preprintnumbers,amsmath,amssymb,floatfix,12pt,prd,superscriptaddress,nofootinbib]{revtex4}
\usepackage{graphicx}
\usepackage{epsfig}
\usepackage{bm}
\usepackage{amsfonts}

\def\ben{\begin{equation}}
\def\een{\end{equation}}

 \def\bd{\begin{document}} \def\ed{\end{document}}
\def\ds{\documentstyle} \let\fr=\frac \let\bl=\bigl \let\br=\bigr
\let\Br=\Bigr \let\Bl=\Bigl
\let\bm=\bibitem
\let\na=\nabla
\let\pa=\partial \let\ov=\overline
\newcommand{\be}{\begin{equation}}
\newcommand{\ee}{\end{equation}}
\def\ba{\begin{array}}
\def\ea{\end{array}}
\def\ft#1#2{{\textstyle{\frac{\scriptstyle #1}{\scriptstyle #2} } }}
\def\fft#1#2{{\frac{#1}{#2}}}
\def\del{\partial}
\def\vp{\varphi}
\def\sst#1{{\scriptscriptstyle #1}}
\def\oneone{\rlap 1\mkern4mu{\rm l}}
\def\td{\tilde}
\def\wtd{\widetilde}
\def\ie{{\it i.e.\ }}
\def\dalemb#1#2{{\vbox{\hrule height .#2pt
        \hbox{\vrule width.#2pt height#1pt \kern#1pt
                \vrule width.#2pt}
        \hrule height.#2pt}}}
\def\square{\mathord{\dalemb{6.8}{7}\hbox{\hskip1pt}}}
\newcommand{\ho}[1]{$\, ^{#1}$}
\newcommand{\hoch}[1]{$\, ^{#1}$}
\newcommand{\bea}{\setlength\arraycolsep{2pt} \begin{eqnarray}}
\newcommand{\eea}{\end{eqnarray}}
\newcommand{\ra}{\rightarrow}
\newcommand{\lra}{\longrightarrow}
\newcommand{\Lra}{\Leftrightarrow}
\newcommand{\bp}{\tilde \beta^\prime}
\newcommand{\tr}{{\rm tr} }
\newcommand{\Tr}{{\rm Tr} }
\def\0{{\sst{(0)}}}
\def\1{{\sst{(1)}}}
\def\2{{\sst{(2)}}}
\def\3{{\sst{(3)}}}
\def\4{{\sst{(4)}}}
\def\5{{\sst{(5)}}}
\def\6{{\sst{(6)}}}
\def\7{{\sst{(7)}}}
\def\8{{\sst{(8)}}}
\def\m{{\sst{(m)}}}
\def\n{{\sst{(n)}}}
\def\cA{{{\cal A}}}
\def\cB{{{\cal B}}}
\def\cF{{{\cal F}}}
\def\cG{{{\cal G}}}
\def\cH{{{\cal H}}}
\def\tV{\widetilde V}
\def\tW{\widetilde W}
\def\tH{\widetilde H}
\def\tE{\widetilde E}
\def\tF{\widetilde F}
\def\tA{\widetilde A}
\def\im{{{\rm i}}}
\def\tY{{{\wtd Y}}}
\def\ep{{\epsilon}}
\def\vep{{\varepsilon}}
\def\bD{{{\bar D}}}
\def\R{{{\mathbb R}}}
\def\C{{{\mathbb C}}}
\def\H{{{\mathbb H}}}
\def\CP{{{\mathbb C}{\mathbb P}}}
\def\RP{{{\mathbb R}{\mathbb P}}}
\def\Z{{{\mathbb Z}}}
\def\bA{{{\mathbb A}}}
\def\bB{{{\mathbb B}}}
\def\bC{{{\mathbb C}}}
\def\bD{{{\mathbb D}}}
\def\bE{{{\mathbb E}}}
\def\bZ{{{\mathbb Z}}}
\def\Re{{{\frak{Re}}}}
\def\Im{{{\frak{Im}}}}
\def\cosec{{\,\hbox{cosec}\,}}
\def\Gm{{\Gamma_{\!\! -}}}
\def\Gp{{\Gamma_{\!\! +}}}
\def\stan{{standard }}
\def\nonstan{{supernumerary }}
\def\p{{\partial}}
\def\kdel#1{{\fft{\del}{\del#1}}}

\def\bog{{Bogomolny }}
\def\om{{\omega}}

\newcommand{\nnr}{\nonumber \\}
\newcommand{\pd}{\partial}
\newcommand{\ud}{\textrm{d}}
\newcommand{\dTH}{T^{\prime \, 0}_\textrm{H}}
\newcommand{\dOi}{\Omega^{\prime \, 0}_i}
\newcommand{\bx}{{\bf x}}
\begin{document}

\title{Holographic superconductors in a model of non-relativistic
gravity}

\author{\textbf{M.R.Setare}}
\email{rezakord@mail.ipm.ir} \affiliation{Department of Campus of
Bijar , University of Kurdistan, Bijar, IRAN}

\author{\textbf{Davood Momeni}\footnote{Corresponding author}}
\email{d.momeni@yahoo.com,d.momeni@tmu.ac.ir}
\affiliation{Department of Physics, Faculty of  Sciences, Tarbiat
Moa'llem University, No. 49, Mofateh Ave., Tehran, Iran}
\author{\textbf{N.Majd}}
\email{naymajd@ut.ac.ir}
 \affiliation{Department of Enginearing
science,University of Tehran, Tehran, IRAN}

\begin{abstract}
\vspace*{1.5cm} \centerline{\bf Abstract} \vspace*{1cm}

We have studied  holographic superconductors with spherical symmetry
in the Ho$\check{\textbf{r}}$ava-Lifshitz gravity by using a semi
analytical method, and also we have calculated the critical
temperature and shown when the condensation will appear in a similar
pattern as in the Einstein-Gauss- Bonnet gravity. We have computed
the dependency of the conductivity as a function of frequency in
this new non-relativistic model of quantum gravity.
\end{abstract}

\maketitle

\newpage

\section{ Introduction}

As a phenomenological fact, superconductivity is usually modeled by
a Landau-Ginzburg Lagrangian where a complex scalar field develops a
condensation in a superconductive phase. To have a scalar
condensation in the boundary theory, Horowitz and his collaborators
\cite{ Horowitz} introduced a U(1) gauge field and a conformally
coupled charged complex scalar field in the black hole
background.That potential corresponding to the
 conformal mass is negative,although above the Breitenlohner-Freedman (BF)
bound \cite{Freedman} it does not cause any instability in the
theory. To solve the negative mass problem Basu et.al and his
collaborators  \cite{Basu}showed that the presence of the vector
potential effectively modifies the mass term of the scalar field as
we move along the radial direction r and allows  the possibility of
developing hairs for the black hole in some parts of the parameter
space. In their model there was  no explicit specification of the
Landau-Ginzburg potential for the complex scalar field. The
development of  condensations relies on a more subtle mechanism
violating the no hair theorem. Further Wen investigated the
holographically dual description of superconductors in (2 + 1)-space
time dimensions in the presence of inhomogeneous magnetic field and
observed that there exist type I and type II superconductor
\cite{arXiv}. the existence of holographic super conductors was
established in\cite{Horowitz,Gubser1}. From the (d dimensional)
field theory point of view, super conductivity is characterized by
condensation of a generally composite charged operator $\hat{O}$ in
low temperatures $T<T_{c}$ .In the gravitationally dual (d+1
dimensional )
 description of the system, the transition to the super
conductivity is observed as a classical instability of a black hole
in an anti-de Sitter (AdS) space against perturbations by a charged
scalar field $\psi $. The instability appears when the black hole
has Hawking temperature $ T=T_{c}$. For lower temperatures the
gravitational dual is a black hole with a non vanishing profile for
the scalar field $\psi $. The AdS/CFT correspondence relates the
quantum dynamics of the boundary operator $\hat{O}$ to a simple
classical dynamics of the bulk scalar field $\psi $
\cite{Gubser2,witten}. Following Hartnoll et al works in
\cite{Horowitz}, and also Maeda and Okamura \cite{Maeda}, we will
find out they studied  the perturbation of the gravitational system
near the critical temperature $T_{c}$, and they obtained the
superconductor's coherence length via AdS/CFT (anti–de
Sitter/conformal field theory) correspondence, and also they added a
small external homogeneous magnetic field to the system, and found a
stationary diamagnetic current proportional to the square of the
order parameter being induced by the magnetic field. Their results
agree with Ginzburg-Landau theory and strongly support the idea that
a superconductor can be described by a charged scalar field on a
black hole via AdS/CFT duality. From a pure classical treatment,
there is more efforts to deal with  BH in Ads backgrounds. Black
holes in anti-de Sitter (AdS) spacetime in several dimensions have
been recently studied. One of the reasons for this intense study is
the AdS/CFT conjecture  stating that there is a correspondence
between string theory in AdS spacetime and a conformal field theory
(CFT) on the boundary of that space. For instance, the M-theory on
$AdS^{4} \times S ^{7}$ is dual to a non-Abelian superconformal
field theory in three dimensions, and type IIB superstring theory on
$AdS^5 \times S^ 5$ seems to be equivalent to a super Yang–Mills
theory
in four dimensions \cite{Maldacena}.\\
Recently, a power-counting renormalizable, ultra-violet (UV)
complete theory of gravity was proposed by Ho\v{r}ava in
\cite{hor2,hor1,hor3,hor4}. Although presenting an infrared (IR)
fixed point, namely General Relativity, in the  UV the theory
possesses a fixed point with an anisotropic, Lifshitz scaling
between time and space of the form $x^{i}\to\ell~x^{i}$,
$t\to\ell^z~t$, where $\ell$, $z$, $x^{i}$ and $t$ are the scaling
factor, dynamical critical exponent, spatial coordinates and
temporal coordinate, respectively. According to the Blas et al
arguments \cite{bla}, it seems that this model must be modified by
some terms to avoid from strong coupling, instabilities, dynamical
inconsistencies and unphysical extra mode.
 As we know that there are two
explicit families of exact solutions for a spherically symmetric
background without projectability condition in HL gravity and other
solutions all are the familiar GR solutions i.e
$Ads^{4}$-Schwarzschild solutions. First solution belongs to the
\cite{KS} known asymptotically flat KS solution  and as we have
shown that in spite of the GR BHs, its timelike geodesics is stable
\cite{sm}. The other non trivial solution was found by Lu-Mei et al.
\cite{Lu}, and recently Tang  \cite{Tang} investigated the general
solutions of the HL theory under both projectability and non
projectability conditions. His paper contains all the former
solutions and at the end of it, he presented  two new families of
exact solutions - only in a neutral case- which both of them are
valid in the corner of the validity of the IR limit of the HL theory
i.e $\lambda=1$ and these solutions can be interpreted as  two new
forms of the BHs for HL gravity.

 Recently the works were done about
the Holographic Superconductors for a new topological BH in HL
gravity describing a topological black hole solution whose horizon
has an arbitrary constant scalar curvature \cite{Cai,bin,Jing}. They
found that it is more applicable for the scalar hair forming, when
the parameter of the detailed balance( $\epsilon$) becomes larger,
and harder when the mass of the scalar field is larger. Also they
calculated the ratio of the gap frequency in the conductivity with
respect to the critical temperature. Briefly they investigated the
effects of the mass of the scalar field and the parameter of the
detailed balance on the scalar condensation, the electrical
conductivity, and the ratio of
 the gap frequency in the conductivity at the critical temperature.\\
 There are many interesting features for critical phenomena and
 superconductivity when we are working on higher orders corrections,
 specially when we are interesting in the  Gauss-Bonnet corrections\cite{betti}.
 The same phenomenology has been discussed by Wang in  series of
 works\cite{bin}. These phenomena and it's physical consequences are very similar
 with our analysis in the HL theory and we can generalize
 their results to our higher order theory in the non relativistic
 regime.

In this work we have discussed a type of solutions which has been
reported in\cite{Lu}. In Sec. 2 we have presented spherically
symmetric black holes' solutions in Ho$\check{\textbf{r}}$ava-
Lifshitz gravity with the action without the condition of the
detailed balance. In Sec. 3 we have explored the scalar condensation
in the Ho$\check{\textbf{r}}$ava-Lifshitz black hole by analytical
approaches. In Sec. 4 the matching solutions and the critical
temperature have been found. In Sec. 5 we have computed the
conductivity of our model and shown the behavior of the real part of
the conductivity as a function of frequency per tempereture. We have
summarized and discussed our conclusions in the last section.

\section{ Solutions of the Ho$\check{\textbf{r}}$ava- Lifshitz
gravity}

 Since in the HL theory, the dynamical quantities are the
shift $N_{i}(t,x)$, lapse $N(t,x)$ and metric $h_{ij}$; therefore in
the ADM formalism \cite{ADM}:
\begin{eqnarray}
ds^2=-N^2dt^2+h_{ij}(dx^{i}+N^{i}dt)(dx^{j}+N^{j}dt)
\end{eqnarray}

If we restricted ourselves to the static metrics $h_{ij}$, there is
two possibility for the  time dependency of the two remaining
functions. In many cases as in Lu-Mei-Pope\cite{Lu}, we can relax
the shift function by a formal going to the Schwarzschild gauge and
rewriting the static solution with spherical symmetry in GR. Thus
for solutions in the usual Schwarzschild gauge the only function is
the lapse. According to the terminology of the Horava theory, a
projectable solution is a solution with a time dependent lapse and a
non projectable one is a vise versa. Many authors consider the non
projectable version as an exact solution. Another problem returns to
the choice of the potential term. The first choice is due to the
detailed balance principle \cite{Sotiriou}, but in the original work
of the Horava in the context of the cosmology this principle implies
a negative cosmological constant in contrary with the observational
evidences. The other problem is avoiding from the ghost excitations
\cite{bla} restricting one to accept a value of the $\lambda
\leq\frac{1}{3}$ or $\lambda>1$. Instability and strong coupling
impose another difficulties for it. Far from all of these problems
we rewrite an explicit spherical symmetric solution for HL theory
following Lu-Mei-Pope  work\cite{Lu}.
\subsection{ New static neutral BH solution}
Following the ADM formalism, the action of this HL gravity  with a
\emph{soft} violation of the \emph{detailed balance} condition is
given by:
\begin{eqnarray}
S&=&\int_{M} dtd^{3}x\sqrt{g}
N(\mathcal{L}_{K}-\mathcal{L}_{V})\\\nonumber
\mathcal{L}_{K}&=&\frac{2}{\kappa^2}\mathcal{O}_{K}
  =\frac{2}{\kappa^2}(K_{ij}K^{ij}-\lambda K^2)\\\nonumber
 \mathcal{L}_{V}&=&\alpha_{6}C_{ij}C^{ij} -
 \alpha_{5}\epsilon_{l}^{ij} R_{im}\nabla_{j}R^{ml} + \alpha_{4}
[R_{ij}R^{ij}- \frac{4\lambda-1}{4(3\lambda-1)} R^2] +\alpha_{2}(R -
3\Lambda_{W})+\frac{\Omega\kappa^{2}\mu^{2}}{8(3\lambda-1)R}\\\nonumber
K_{ij}&=&\frac{1}{2N}(\dot{g_{ij}}-\nabla_{i}N_{j}-\nabla_{j}N_{i})
\end{eqnarray}
The $\alpha_{i}$ are the coupling parameters \cite{Lu}, and $C_{ij}$
is the Cotton tensor \cite{hor3}. With the metric ansatz as in
\cite{Lu}:
\begin{eqnarray}
ds^{2}=-N(r)^{2}dt^{2}+\frac{1}{f(r)}(dr+N^{r}dt)^{2}+r^{2}d\Omega^{2}
\end{eqnarray}
The following solution in the UV region has been found \cite{Lu}:
\begin{eqnarray}
N^{r}=0\\\nonumber
\delta=\frac{2\lambda\pm\sqrt{6\lambda-2}}{\lambda-1}\\\nonumber
\gamma=\delta-1\\
f(r)\equiv f=1-\frac{\Lambda_{W}}{2}r^{2}-\alpha r^{\delta}\\
N(r)\equiv N=\beta r^{-\gamma}\sqrt{f}
\end{eqnarray}
where $\alpha,\beta$ are constants. This solution  is asymptotically
$AdS^{4}$ and thus  it is useful in the AdS/CFT correspondence
scenario for the Holographic superconductivity. The Hawking
temperature is given by the usual Gibbons-Hawking calculus
\cite{gib}, therefore the Unruh temperature can be written in the
form \cite{Konoplya}:
\begin{eqnarray}
T=\frac{N'\sqrt{f}}{2\pi}|_{r=r_{H}}=\frac{\beta}{4\pi}h^{-\gamma}f'(h)=-\frac{\beta}{4\pi}(\Lambda_{W}h+\alpha\delta
h^{\delta-1})
\end{eqnarray}
in order to satisfy  the   positivity of the temperature, we must
require $\beta<0$ when both $\Lambda_{W}$ and $\alpha$ are positive
simultaneously.
\section{Field equations for scalar condensation scenario}
Following the Hartnoll, Herzog and Horwitz general framework to the
holographic superconductors \cite{Horowitz, Horowitz2}, in the limit
where the scalar field does not back-react on the geometry the
solution for the background geometry is that of the dyonic black
hole \cite{rom}. In this paper, the charge density of the background
\cite{Horowitz, Horowitz2,Lu} is neutral, so both the electric and
magnetic charge of the dyonic black hole have been set to zero. The
Maxwell-scalar sector is decoupled from the gravity sector,
therefore the minimal ingredients we need to describe a holographic
superconductor are conserved energy momentum $ T^{\mu\nu}$, Global
U(1) symmetry, conserved current $J^{\mu}$ and finally charged
operator $ \hat{O}$ condensing at low temperature ($\mu$, $\nu $
runs over $t$, $x$, $y$). The most basic entries in the AdS/CFT
dictionary \cite{Gubser2, witten} tell us that there is a mapping
between field theory operators and fields in the bulk . In
particular, $ T^{\mu\nu}$ will be dual to the bulk metric $g_{ab}$,
the current $J^{\mu}$ will be dual to a Maxwell field in the bulk
$A_{a}$, and the dual of charged scalar field $\psi $ is $\hat{O}$ (
here $a$, $b$ runs over $t$, $x$, $y$, $r$). We can now study the
Maxwell-scalar theory in the black hole background with Lagrangian:
\begin{eqnarray}
\textbf{L}=-\frac{1}{4}F^{2}-|\partial\psi-iA\psi|^{2}+2\frac{\bar{\psi}\psi}{L^{2}}
\end{eqnarray}
The only  dimensional parameter in the Lagrangian is L related to
the AdS radius, and
 the full set of equations of motion for
the fields $\psi$ and $A_{\mu}$ are :
\begin{eqnarray}
\frac{1}{\sqrt{-g}}\partial_{\mu}(\sqrt{-g}g^{\mu\nu}(\partial_{\nu}\psi-iA_{\nu}\psi))+\frac{2}{L^{2}}\psi-ig^{\mu\nu}A_{\mu}(\partial_{\nu}\psi-iA_{\nu}\psi)=0\\
\frac{1}{\sqrt{-g}}\partial_{\nu}(\sqrt{-g}g^{\nu\lambda}g^{\mu\sigma}F_{\lambda\sigma})-
g^{\mu\nu}(i(\bar{\psi}\partial_{\nu}\psi-\partial_{\nu}\bar{\psi}\psi)+2A_{\nu}\bar{\psi}\psi)=0
\end{eqnarray}
respectively, and we can have the same equation for $\bar{\psi}$ by
complex conjugating of equation (9). We take the ansatz:
\begin{eqnarray}
\psi=\psi(r),A_{t}=\phi(r),A_{a}=0,a={r,\theta,\phi}
\end{eqnarray}
It is then  suitable to take the phase of $\psi$  to be constant.
All other fields are set to be zero. Under this ansatz, the
equations of motion simplify to:
\begin{eqnarray}
r^{\gamma-2}(r^{2-\gamma}f\psi')'+\frac{2}{L^{2}}\psi+N^{-2}\phi^{2}\psi=0\\
r^{\gamma-2}(r^{2+\gamma}\phi')'-2\phi\psi^{2}r^{2\gamma}f^{-1}=0
\end{eqnarray}
where a prime denotes the derivative with respect to r, and we have
to notify that if $\gamma=0$, these equations will reduce to the
 ones in \cite{Horowitz,Horowitz2,tam}. We define a mass parameter
as:
\begin{eqnarray}\nonumber
m^{2}L^{2}=-2
\end{eqnarray}
The field equations  (12), (13) can  be written as the next set:
\begin{eqnarray}
\psi''+(\frac{2-\gamma}{r}+\frac{f'}{f})\psi'+(\frac{r^{2\gamma}}{\beta^{2}f^{2}}\phi^{2}-\frac{m^{2}}{f})\psi=0\\
\phi''+(2+\gamma)r^{2\gamma-1}\phi'-2\phi\psi^{2}r^{2\gamma}f^{-1}=0
\end{eqnarray}
If $\beta=1,\gamma=0$ we recover again the results of
\cite{Horowitz,Horowitz2,cai}. We must note an important fact about
the limiting process  to achieve the Lu et al solution given in
\cite{cai}. The limiting process $\gamma\rightarrow0$ is valid for
both different values of the $\lambda=1,3>1$. The Lu et al solution
recovers both of these values, although we observe from the form of
the lapse function
 that these values lead to the same metric functions .

 Examining these fields
equations at the horizon and assuming that the scalar field must be
regular on the horizon, we can observe that we have the next set of
the auxiliary boundary conditions:
\begin{eqnarray}
\psi'_{r_{H}}=\frac{m^{2}}{f'_{h}}\psi_{r_{H}}\\
\phi_{r_{H}}=0
\end{eqnarray}
in which $r_{H}$ is the horizon radius of the black hole, i.e. the
largest root of $f(r) = 0$.

\subsection{ Solving the general equations in the asymptotic
region }
 In the vicinity of the black hole, Eqs (14), (15) can be solved by
making a change of variable, $r\rightarrow r_{H}$, and  setting the
radius of $AdS^{4}$ to be L = 1 \cite{cai}. In \cite{cai} also the
case $m^{2}=0$ was discussed both via numerical and semi analytical
methods. In this manuscript we limited ourselves  only to this
special case $m^{2}=0$. We can easily guess their behavior in the
large r limit. In order to find the asymptotic behavior of the field
we must determine when in the IR region $\lambda>1$, the exponent
$\delta$ is positive or negative. There are two different kinds of
the exponent $\delta$ which we denote them by $\delta_{+},
\delta_{-}$. We mention here that for a sufficient large value of
the   $\lambda$ the value of the exponent $\delta_{-}$ remains below
2. Thus for all values of the $\lambda>1$, we have the next limiting
values:
\begin{eqnarray}
\lim_{\lambda\rightarrow 1^{+}}(\delta_{+})&=&+\infty\\
\lim_{\lambda\rightarrow 1^{+}}(\delta_{-})&=&\frac{1}{2}\\
\frac{1}{2}<\delta_{-}&<&2\\
2<\delta_{+}&<&\infty\\
1<\gamma_{+}&<&\infty\\
-\frac{1}{2}<\gamma_{-}&<&1
\end{eqnarray}
\subsection{Approximation techniques}
According to the method discussed in \cite{kan} we must find the
approximate solutions near the horizon, then generalize it to the
asymptotic AdS region and smoothly match the solutions at an
intermediate point. By introducing a new radial-like coordinate as:
\begin{eqnarray}
\xi=\frac{r_{H}}{r}
\end{eqnarray}
we can rewrite the equations (14), (15) in terms of the new
coordinate $\xi$\footnote{We limited ourselves to a massless case
$m^{2}=0$}:
\begin{eqnarray}
\ddot{\psi}+(\frac{\gamma}{\xi}+\frac{\dot{f}}{f})\dot{\psi}+(\frac{r_{H}^{2\gamma+4}\xi^{-2\gamma-4}}{\beta^{2}f^{2}}\phi^{2})\psi=0\\
\ddot{\phi}+(\frac{2}{\xi}-r_{H}^{2\gamma}(2+\gamma)\xi^{-1-2\gamma})\dot{\phi}-2\psi^{2}r_{H}^{2\gamma+2}\xi^{-2\gamma-4}f^{-1}\phi=0
\end{eqnarray}
where a dot now denotes $\frac{d}{d\xi}$ and we observe that for the
interval out of the horizon this coordinate smoothly covers all
points of the strip:
\begin{eqnarray}
r_{H}<r<\infty , \hspace{0.5cm} 0<\xi<1
\end{eqnarray}
The boundary conditions (16) and (17) in the massless limit with the
regularity at the horizon  $\xi = 1$ become:
\begin{eqnarray}
\phi(1)=0,\hspace{0.5cm} \dot{\psi}(1)=0
\end{eqnarray}
With this change of the variable the equations (14) and (15) convert
to the next set (16) and (17), which must be solve near horizon i.e
$\xi=1$ with auxiliary boundary conditions (28). Our main goal is to
find the coefficients and powers in (25), (26) and also matching
these two solution in an intermediate point.

\subsection{Solutions near the horizon:$\xi=1$}
We can expand $\psi(r)$ and $\phi(r)$  in a Taylor series near the
horizon as:
\begin{eqnarray}
\phi(\xi)=\phi(1)-\dot{\phi}(1)(1-\xi)+\frac{1}{2}\ddot{\phi}(1)(1-\xi)^{2}+...\\
\psi(\xi)=\psi(1)-\dot{\psi}(1)(1-\xi)+\frac{1}{2}\ddot{\psi}(1)(1-\xi)^{2}+...
\end{eqnarray}
According to the equation (28), for a massless scalar field, we have
$\dot{\psi}(1) = 0 $ and $\phi(1) = 0 $, and without loss of
generality we take $\dot{\phi}(1)<0,  \psi(1)>0$ to have $\phi(1)$
and $\psi(1)$ positive. Expanding (26) near $\xi = 1$ gives:
\begin{eqnarray}
\ddot{\phi}(1)=(\frac{2\psi(1)^{2}}{\dot{f}(1)}r_{H}^{2\gamma+2}+r_{H}^{2\gamma}(2+\gamma)-2)\dot{\phi}(1)
\end{eqnarray}
Thus, we get the approximate solution:
\begin{eqnarray}
\phi(\xi)=\dot{\phi}(1)(-(1-\xi)+\frac{1}{2}(1-\xi)^{2}(\frac{2\psi(1)^{2}r_{H}^{2\gamma+2}}{\dot{f}(1)}+r_{H}^{2\gamma}(2+\gamma)-2))
\end{eqnarray}
Similarly, from (25), the 2'nd order coefficients of $\psi$  can be
calculated as:
\begin{eqnarray}
\ddot{\psi}(1)=-\frac{r_{H}^{2\gamma+4}}{2\beta^{2}}\psi(1)(\frac{\dot{\phi}(1)}{\dot{f}(1)})^{2}
\end{eqnarray}
where we used Hopital rule at the second term, therefore an
approximate solution near the horizon is:
\begin{eqnarray}
\psi(\xi)=\psi(1)(1-\frac{r_{H}^{2\gamma+4}}{4\beta^{2}}(\frac{\dot{\phi}(1)}{\dot{f}(1)})^{2})(1-\xi)^{2}
\end{eqnarray}

\subsection{Solutions in the asymptotic AdS region}

 In the asymptotic AdS region $\xi=0$, the solutions are:
\begin{eqnarray}
\psi=D_{+}\xi^{\lambda_{+}}+D_{-}\xi^{\lambda_{-}}\\
\phi=\mu-q\xi
\end{eqnarray}

where $\mu$ is the chemical potential and $q$ is the charge density
on the boundary\footnote{Our  compendium follows what mentioned in
 the Gregory et.al work\cite{Gregory}} . At the boundary of a
(2+1)-dimensional field theory, $\mu$ is of mass dimension one and
$q=\rho/r_H$ is of mass dimension two. From the boundary behaviors,
we can read off the expectation value of operator $\hat{O}$ dual to
the field. From \cite{Horowitz,Horowitz2,kel}, we know that , both
of these falloffs are normalizable, and in order to keep the theory
stable \cite{Horowitz}, we should impose  the following equations:
\begin{eqnarray}
D_{+}=0, \hspace{0.5cm}<\hat{O}_{-}>=\sqrt{2}D_{-}\\
D_{-}=0,\hspace{0.5cm}<\hat{O}_{+}>=\sqrt{2}D_{+}
\end{eqnarray}
where the   factor  $\sqrt{2}$  is a convenient
normalization\cite{Horowitz}. The index $i$ in  $D_{i}$ represents
the scaling dimension $\lambda_{O}$ of its dual operator
$<\hat{O}_{i}>$, i.e. $\lambda_{O_{i}}=i$. Note that these are not
entirely free parameters, as there is a scaling degree of freedom in
the equations of motion. As in [1], we impose that $\rho$ is fixed,
which determines the scale of this system. For $\psi$, both of these
falloffs are normalizable, so we can impose the condition either
$D_{-}$ or $D_{+}$ vanish. We take $D_{-} = 0$, for simplicity.
 Now  we must find the solutions of
the equations (25) and (26) with the boundary conditions mentioned
above. Since the dimension of temperature T is of mass dimension
one, the ratio $T^{2}/\rho$ is dimensionless. Therefore increasing
$\rho$, while $T$ is fixed, is equivalent to decrease T while $\rho$
is fixed. We must show that when $\rho > \rho_{c}$, the operator
condensate will appear; this means when $T < T_{c}$ , there will be
an operator condensation, that is to say the superconducting phase
occurs.\\
We limited ourselves only to the case $\delta_{+}>2,\gamma_{+}>1$.
Remembering for a general 2'nd order differential equation, we can
write (25) in the following self-adjoint form:
\begin{eqnarray}
\ddot{\Psi}+P(x)\dot{\Psi}+Q(x)\Psi=0
\end{eqnarray}
The change of the variable $\Psi(x)=e^{-1/2\int P(x)dx}\Xi(x)$
converts it to the next Schrodinger like equation:
\begin{eqnarray}
\ddot{\Xi}(x)+(-1/2\dot{P}-1/4P^{2}+Q)\Xi(x)=0
\end{eqnarray}
For (25) we have:
\begin{equation}
P=\frac{\gamma}{\xi}+\frac{\dot{f}}{f},\hspace{0.5cm}Q=\frac{h^{2\gamma+4}\xi^{-2\gamma-4}}{\beta^{2}f^{2}}\phi^{2},
\hspace{0.5cm}\Psi=\frac{\Xi(x)}{\sqrt{f}\xi^{\gamma/2}}\\
\end{equation}
In AdS asymptotic region with the metric function $f\sim -\alpha
\xi^{-\delta}$, the field equation (40) is converted to the:
\begin{equation}
\xi^{2}\ddot{\Xi}(\xi)+\eta\Xi(\xi)=0,\hspace{0.5cm}\eta=1/4(1-(\gamma-\delta-1)^{2})=-\frac{3}{4}
\end{equation}
This is a standard Euler-Cauchy equation which has the following
exact solution:
\begin{eqnarray}
\Xi(\xi)=\Xi_{+}\xi^{m_{+}}+\Xi_{-}\xi^{m_{-}},\hspace{0.5cm}m_{\pm}=\frac{3}{2},-\frac{1}{2}\\
\psi(\xi)=D_{+}\xi^{2}+D_{-}\\
\phi(\xi)=\mu-\rho\xi
\end{eqnarray}
The new set of coefficients $ D_{\pm}$ are some functions of  the
$\Xi_{\pm},\alpha,...$.

\section{Matching and phase transition}
Now we will match the solutions (32),(34), and (44), (45) at
$\xi_{m}$. Allowing $\xi_{m}$ to be arbitrary does not change
qualitative features of the analytic approximation, and  more
importantly, it does not give a big difference in numerical values;
therefore for simplicity in demonstrating our argument we will take
$\xi_{m}=1/2$. In order to connect our two asymptotic solutions
smoothly, we require continuity in our fields and their first
derivatives at the crossing point $\xi_{m}=1/2$, therefore following
 four conditions should be satisfied\footnote{We have set $\psi(1) = a $
and $-\dot{\phi}(1) = b$, $(a, b > 0)$ for clarity,
$\dot{f}(1)=-\frac{4\pi T}{\beta}h^{\gamma+1}$}:
\begin{eqnarray}
D_{-}+\frac{D_{+}}{4}=a(1-\frac{b^{2}r_{H}^{2}}{256 \pi^{2}T^{2}})\\
D_{+}=\frac{ab^{2}r_{H}^{2}}{64\pi^{2} T^{2}}\\
\mu-\frac{\rho}{2}=-b(-\frac{3}{4}+\frac{1}{8}(-\frac{a^{2}\beta
r_{H}^{\gamma+3}}{2\pi T}+r_{H}^{2\gamma}(2+\gamma)))\\
\rho=b(2-\frac{1}{2}(-\frac{a^{2}\beta r_{H}^{\gamma+3}}{2\pi
T}+r_{H}^{2\gamma}(2+\gamma)))
\end{eqnarray}

after setting $D_{-}=0$, we obtain from equations (46) and (47):
\begin{eqnarray}
D_{+}=2a&=&\frac{ab^{2}r_{H}^{2}}{64\pi^{2} T^{2}}\\
b&=&\frac{8\sqrt{2}\pi T}{r_{H}}\\
b&=&\tilde{b}T
\end{eqnarray}
where ($\tilde{b}:=\frac{8\sqrt{2}\pi}{r_{H}}$) and also from
equations (48) and (49)we have:
\begin{eqnarray}
a^{2}=\frac{16\pi T}{b\beta
h^{\gamma+3}}[\mu-\frac{\rho}{2}-\frac{3}{4}b(1-\frac{h^{2\gamma}(2+\gamma)}{6})]\\
a^{2}=\frac{4\pi T}{b\beta
h^{\gamma+3}}[\rho-2b(1-\frac{h^{2\gamma}(2+\gamma)}{4})]
\end{eqnarray}
where( $h:=r_{H}$) and then we conclude that:
\begin{eqnarray}
b=4\mu-3\rho
 \end{eqnarray}
and  we can define the critical point, $T_{C}$ as:
\begin{eqnarray}
T_{C}=\frac{\rho}{2\tilde{b}(1-\frac{h^{2\gamma}(2+\gamma)}{4})}
\end{eqnarray}
\begin{figure}
\centering
 \includegraphics[width=10cm,angle=270] {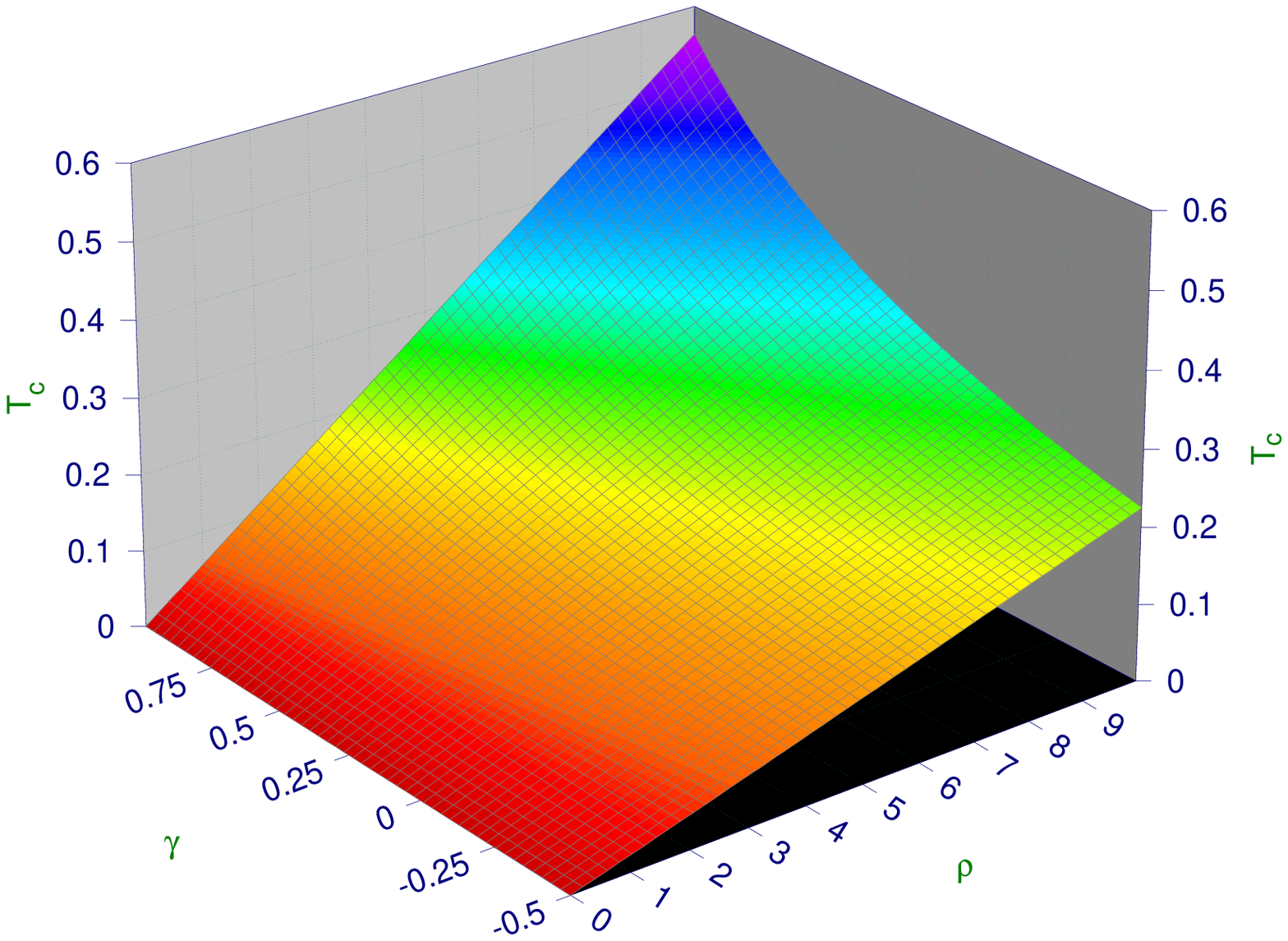}
 \caption{ A plot of the critical temperature as a function of $\rho$
and $\gamma$ varying in the range of $-0.5<\gamma<1$}
  \label{1.eps}
\end{figure}

Figure (1) shows the the dependence of $T_{c}$ as a function of
$\rho$ and $\gamma$. As we see when $\rho=0$ for different values of
$\gamma$ the value of $T_{c}$ is equal to zero, and in the case
$\gamma=0$ there is a linear dependency of $T_{c}$ with respect to
the varying parameter $\rho$. This is also mentioned in the figure
(2) . As we see in the  figure(2) when $\rho=0$ the magnitude of
$T_{c}$ is equal to zero and when $\rho$ goes higher the  $T_c$ also
goes higher with linear dependency. In the figure (3) we show the
dependency of $T_{c}$ with respect to $\gamma$ in the range of
$-0.5<\gamma<1$, when $\rho$ is fixed (for example in that case
$\rho=10$). With increasing of $\gamma$, the values of $T_{c}$ also
increase but not linearity.

\begin{figure}
\centering
 \includegraphics[width=10cm,angle=270] {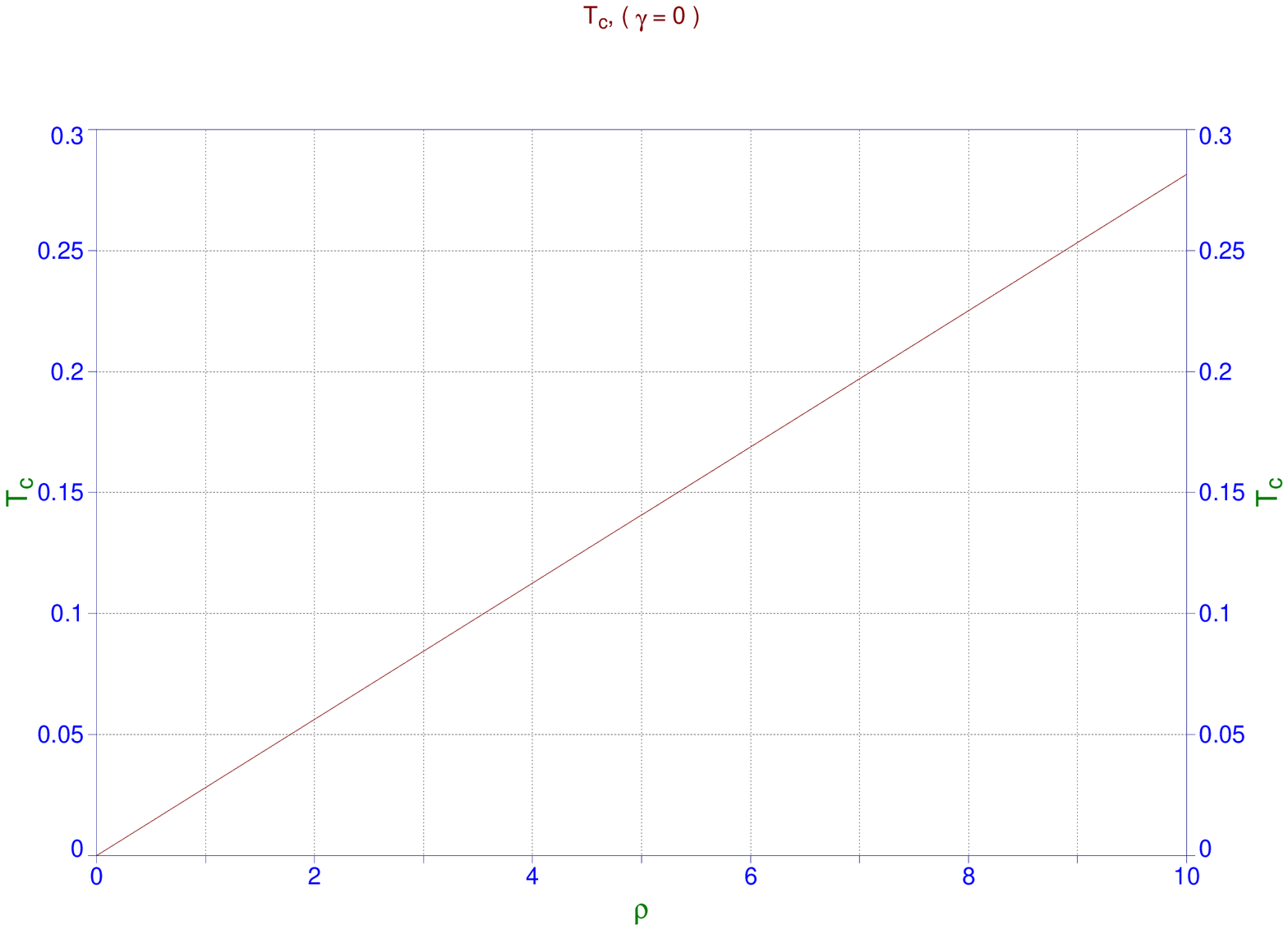}
  \caption{ A plot of the $T_c$ as a function of $\rho$ ( $\gamma=0$
). In  this plot  $\gamma=0$ as we change $\rho$ between 0 and 10.
As we see there is linear dependency with respect to parameter
$\rho$.}
  \label{2.eps}
\end{figure}

\begin{figure}
\centering
 \includegraphics[width=10cm,angle=270] {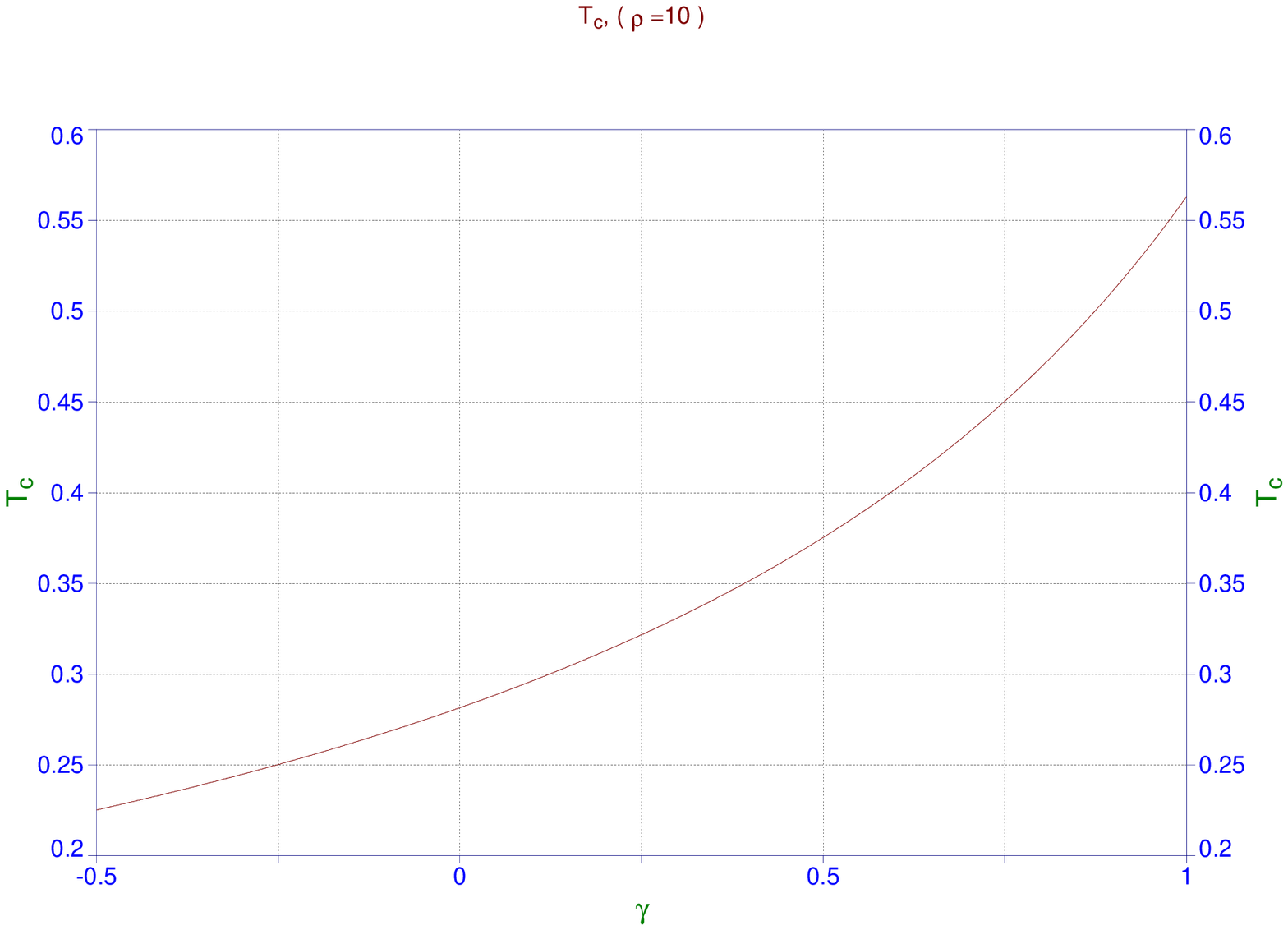}
  \caption{ A plot of the $T_c$ as a function of $\gamma$ ( $\rho=10$ ). This plot  shows how by varying
$\gamma$ the values of $T_c$ change when $\rho$ is fixed ( for this
case $\rho=10$ ). }
  \label{3.eps}
\end{figure}

Noting that in order to remain the temperature $T_{C}$ positive, we
must have $h^{\gamma}<\frac{2}{\sqrt{2+\gamma}}$ , and according to
the equation(7) we can conclude
that($\frac{3}{8}<h<\frac{2}{\sqrt{3}}$) , and it could be
reasonable to choose $h=1$ . Near the critical temperature the
AdS/CFT dictionary gives the relation below :
\begin{eqnarray}
<\hat{O}_{+}>=\sqrt{2}
D_{+}=2\sqrt{2}a=4\sqrt{\frac{2\pi\rho}{\tilde{b}\beta
h^{\gamma+3}}}(1-\frac{T}{T_{C}})^{1/2}
\end{eqnarray}

We observe that $<\hat{O}_{+}>$ is zero at $T = T_{C}$, the critical
point, and condensation occurs for $T < T_{C}$. The continuity of
the transition can be checked by computing the free
energy\cite{Horowitz}. We also see a behavior $<\hat{O}_{+}>\propto
(T_{C}-T)^{1/2}$ which is a typical mean field theory result for a
second order phase transition\cite{Gregory}.

Figure (4) shows $\langle {\hat O}_+ \rangle$ as a function of
temperature normalizing by $T_{c}$ for a variety of values of $\rho$
and $\beta$. Each line in the plot forms the characteristic curve of
$\langle {\hat O}_+ \rangle$ condensing at some critical
temperature. For simplicity we chose five values of $\rho$ and
$\beta$ to display the features of the system and showing how
varying $\beta$ and $\rho$ effect the height of $\langle {\hat O}_+
\rangle$. In this figure according to the equation (7), in order to
have positive Unruh temperature we must require $\Lambda_{W}<0$.

\begin{figure}
\centering
\includegraphics[width=10cm,angle=270]{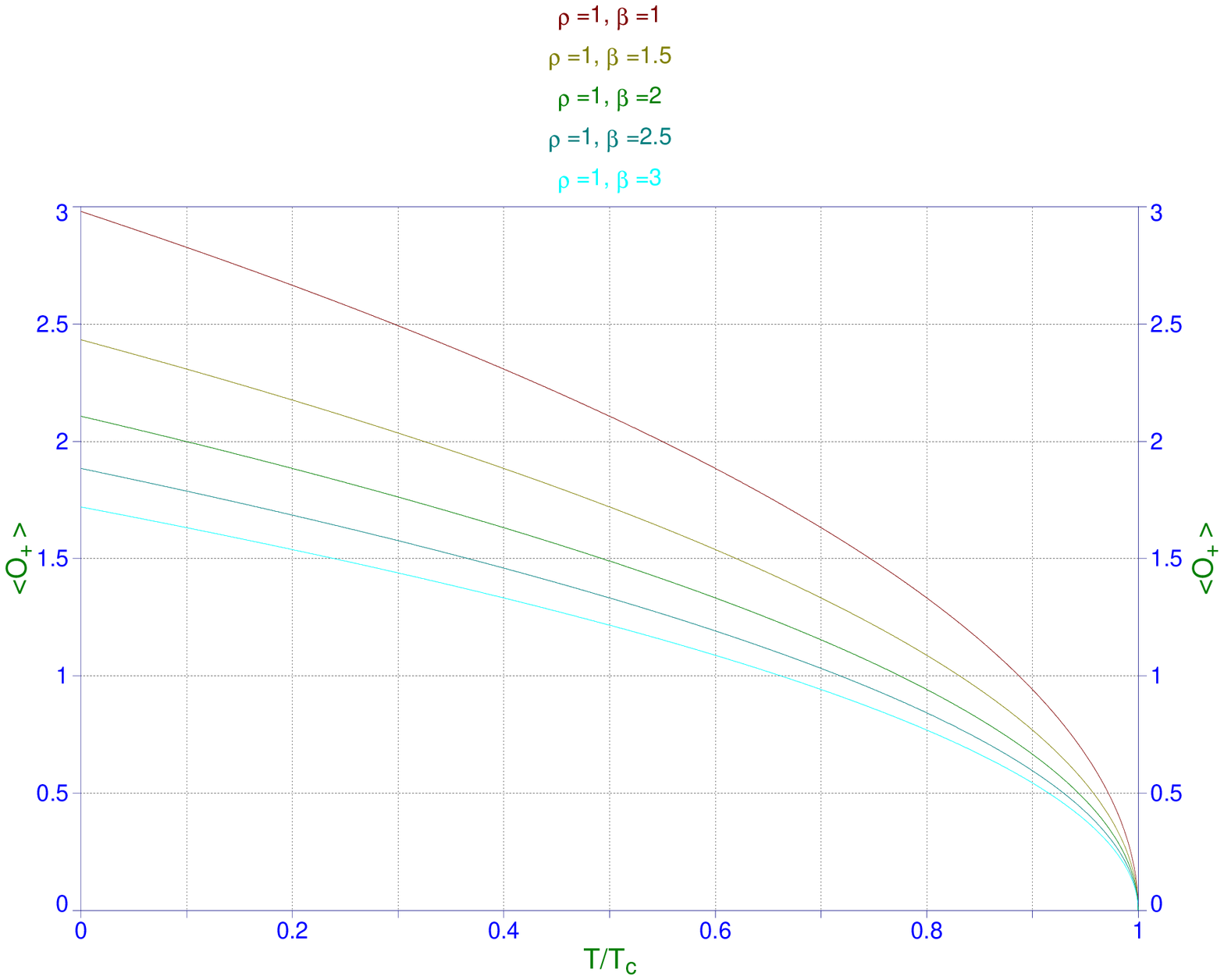}
  \caption{ The plot of the condensation as a function of
$\frac{T}{T_c}$ for a selection of values of $\rho$ and $\beta$. In
this plot the value of $\rho$ is fixed ( in this case is equal to
one) and the value of $\beta$ from top to down is equal to 1, 1.5,
2, 2.5, 3. As we see the height of $\langle {\hat O}_+ \rangle$ is
decreasing as the values of $\beta$ increase, and the condensation
occurs when $ T < T_c$. }
  \label{4.eps}
\end{figure}

\begin{figure}
\centering
 \includegraphics[width=10cm,angle=270] {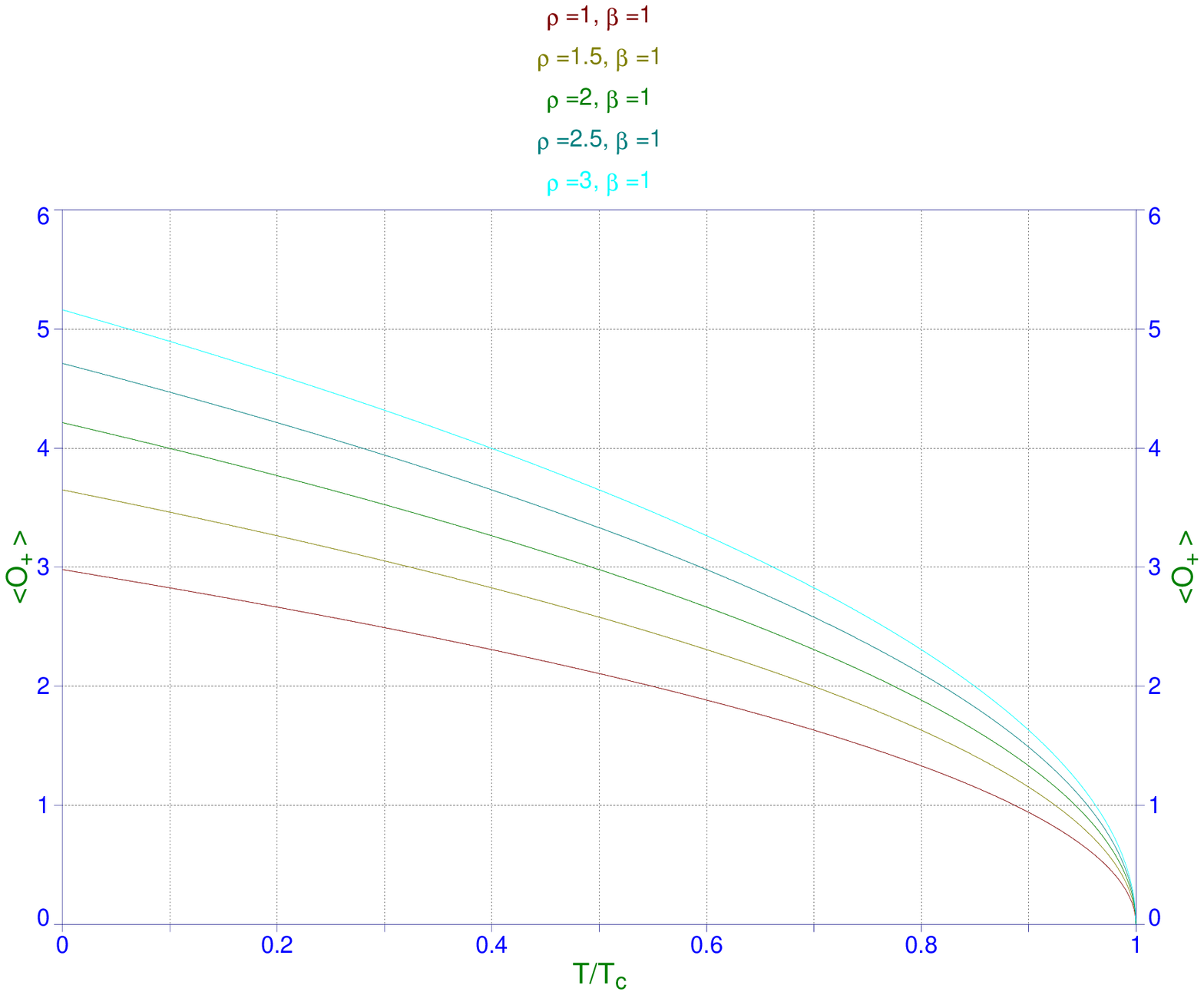} 
  \caption{ The plot of the condensation as a function of
$\frac{T}{T_c}$ for a selection of values of $\rho$ and $\beta$. In
this plot the value of $\beta$ is fixed ( in this case is equal to
one) and the value of $\rho$ from down to top is equal to 1, 1.5, 2,
2.5, 3. As we see the height of $\langle {\hat O}_+ \rangle$ is
increasing as the values of $\beta$ increase, and the condensation
occurs when $ T < T_c$. }
  \label{5.eps}
\end{figure}

 As we see  in figure (4) increasing
$\beta$ reduces the value of $\langle {\hat O}_+ \rangle$. We also
see that the condensation appears when $T=T_c$.

 Figure (5) shows that the  effect of  increasing $\rho$ is to increase the
height of these graphs ( $\langle {\hat O}_+ \rangle$ ), in similar
way mentioned in figure (4), the condensation happens at $T=T_c$.

\section{ Conductivity}
In order to  compute the electric conductivity in dual CFT , we must
solve the Maxwell equation for the fluctuations of the vector
potential $A_{x}(r,t)$, located in the bulk. We assume that the time
dependence of the field is $e^{-i\omega t}$ and then the field
equation of this component reads as:
\begin{equation}
   A^{''}+\left(\frac{f^{'}}{f}-\frac{(\gamma+1)}{r}\right)A^{'}+
   \left[\frac{r^{2\gamma}\omega^{2}}{f^{2}\beta^{2}}-\frac{2\psi^{2}}{f}\right]A=0
 \end{equation}
which is what mentioned in the papers
\cite{Gregory,Gregory2,Gregory3} in the special case $\kappa=0$ and
$\alpha=\frac{L^{2}}{4}$ where from the metric ansatz we have
concluded that
 $e^{\nu}=\beta r^{-\gamma}$. The causal behavior is obtained with
 imposing an ingoing wave boundary condition at the horizon
 \cite{bc}. The desired asymptotic behavior of the Maxwell field at
 large distance is

\begin{equation}
   A_{x}=A_{x}^{(0)}+\frac{A_{x}^{(1)}}{r}
 \end{equation}
According to the AdS/CFT dictionary, the dual source and expectation
value for the current are given by

\begin{equation}
   A_{x}=A_{x}^{(0)},<J_x>=A_{x}^{(1)}
 \end{equation}

Now using Ohm's law we can obtain the conductivity as

\begin{equation}
   \sigma(\omega)=\frac{-i A_{x}^{(1)}}{\omega A_{x}^{(0)}}
 \end{equation}
Thus we must solve (58) numerically and obtain the imaginary part of
the conductivity $\sigma(\omega)$ for a set of parameters
$\Lambda_W=-2,\beta=-1,\delta=-1$.  There is a delta function at
$\omega=0$ which appears as
 $ T<T_{C}$, and from the Kramers-Kronig relation we can see that the
 real part of the conductivity contains a delta function and the
 imaginary part has a simple pole at $\omega=0$. Thus
 the superfluid density is of the delta function \cite{Horowitz}
\begin{equation}
   Re(\sigma(\omega))\sim \pi \delta (\omega)
 \end{equation}
The  figure(6) shows the behavior of the real part of conductivity
as a function of frequency per temperature for different values of
the HL parameter $\alpha$.

\begin{figure}
\centering
 \includegraphics[width=10cm,angle=0] {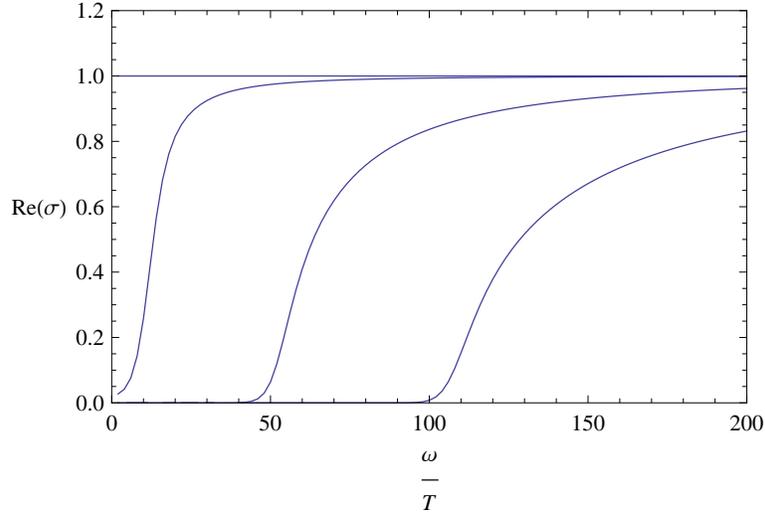} 
  \caption{ The real part of conductivity $ Re(\sigma(\omega))$ as a function of the frequency $\frac{\omega}{T}$ for $O_1$ operator}
  \label{6.eps}
\end{figure}

 We can solve (58) for operator $O_2$ for the former set of the
parameters. The result graph has been shown in the figure(7).
\begin{figure}
\centering
 \includegraphics[width=10cm,angle=0] {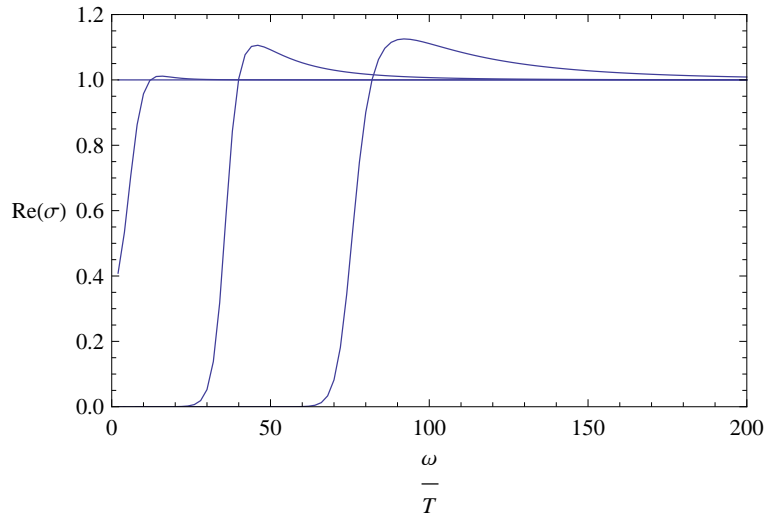} 
  \caption{ The real part of conductivity $ Re(\sigma(\omega))$ as a function of the frequency $\frac{\omega}{T}$ for $O_2$ operator}
  \label{7.eps}
\end{figure}

\section{Conclusion}
In the present work, we have built a holographic model for a
non-relativistic system showing superconductivity. We have used a
black hole background which comes from the
Ho$\check{\textbf{r}}$ava-Lifshitz gravity, and we have studied
analytically, holographic superconductors in this new kind of the
asymptotic AdS solutions. We also have analytically  solved the
system in the probe limits, near horizon and asymptotic region. We
have found that there is also a critical temperature like the
relativistic case, below which a charged condensation field appears
by a second order phase transition, and also we have found out below
a critical temperature $ T_{C}$, the condensation field appears and
obtains finite value. We can conclude that  as the condensation
field becomes heavier, the transition happens more observable. Also
 the conductivity has been computed and the variation of the critical temperature and
conductivity  with respect to the parameters of the metric function
have been shown. We numerically obtain the conductivity as a
function of the frequency for a wide range of the parameters. We
show that the Gauss-Bonnet theory in five dimension and
Ho$\check{\textbf{r}}$ava-Lifshitz theory in critical exponent $z=3$
and in four dimension share some similar features.
\section{Acknowledgment}
The authors would like to thank  Bin Wang
 from INPAC (China) for recommending useful
references in Gauss-Bonnet superconductors and proposing excellent
observations and helpful suggestions  resulted in substantial
improvements of the presentation and outcomes. Also we thank Betti
Hartmann from Jacobs university for suggesting former references for
Gauss Bonnet superconductors.

\end{document}